# Near-field radiative heat transfer between nanostructures in the deep sub-wavelength regime

Raphael St-Gelais,[1,2] Linxiao Zhu,[3] Shanhui Fan,[3] and Michal Lipson*[,1,2]

[1]School of Electrical and Computer Engineering, Cornell University, Ithaca, New York 14853, United States
[2]Department of Electrical Engineering, Columbia University, New York, New York 10027, United States
[3]Ginzton Laboratory, Stanford University, Stanford, California 94305, United States
* e-mail: ml3745@columbia.edu

**Introductory paragraph**

Radiative heat transfer between parallel objects separated by deep sub-wavelength distances and subject to large thermal gradients (>100 K) could enable breakthrough technologies for electricity generation[1-5] and thermal transport control[6-8]. However, thermal transport in this regime has never been achieved experimentally due to the difficulty of maintaining large thermal gradients over nm-scale distances while avoiding other heat transfer mechanism such as conduction. Previous experimental measurement between parallel planes [1,9-13] were limited to distances greater than 500 nm[11] (with a 20 K thermal gradient), which is much larger than the theoretically predicted distance (<100 nm) required for most applications[1-8]. Here we show near-field radiative heat transfer between parallel nanostructures in the deep sub-wavelength regime using high precision micro electromechanical (MEMS) displacement control. We also exploit the high mechanical stability of structures under high tensile stress to minimize thermal buckling effects and maintain small separations at large thermal gradients. We achieve an enhancement of heat transfer of almost two orders of magnitude relative to the far-field limit, which corresponds to a 54 nm separation. We also achieve a high temperature gradient (260 K) between the cold and hot surfaces while maintaining a ~100 nm distance.

**Main text**

Radiative heat transfer between objects separated by deep sub-wavelength distances can exceed the conventional laws of thermal radiation[14,15], while being concentrated over a quasi-monochromatic frequency range[16]. This heat transfer occurs through evanescent coupling of thermally excited surface resonances, and consequently scales inversely with separation (as $1/d^\alpha$, where $\alpha$ is a geometry dependent factor). The separation ($d$) at which this regime occurs depends on the geometry of the system and on the materials involved, but it is generally around 200 nm for identical parallel structures relying on surface phonon resonances[17]. For applications such as electricity generation and thermal control, the two materials involved are non-identical, such that resonant coupling is less efficient and separations <100 nm are typically required[3,18].

The unique features of heat transfer in the deep sub-wavelength regime (i.e., large magnitude and quasi-monochromatic spectral distribution) could allow high efficiency generation of electricity from heat[1,3-5] and novel thermal control devices[6,8]. For example, near-field heat transfer between a hot thermal emitter and a cold photovoltaic cell could allow energy generation with a greater efficiency (>30% for a $T$ = 600 K heat source[5]) than using either thermoelectric generator or single junction solar photovoltaic cells. It is also predicted that engineering the surface resonance frequencies of parallel structures could allow novel thermal control devices such as thermal rectifiers[6,7] or thermal transistors[8].

These applications rely on radiative heat transfer between parallel structures in the deep sub-wavelength regime, and on a high temperature gradient between them, neither of which have been achieved experimentally. Near-field enhancement of heat transfer was demonstrated between parallel plates using active parallelism control[10,12,13], or mechanical spacers[1,9,11]. The smallest separation achieved in these experiments is 500 nm[11] (with a 20 K thermal gradient), which is small enough to overcome the far-field blackbody radiation limit, but not enough to reach the deep sub-wavelength regime where the heat flux is effectively concentrated around a single frequency[17]. Furthermore, the highest temperature gradient ($\Delta T$) achieved in these experiments is 85 K[9] (for a 1.6 µm separation), which is relatively small, especially for energy generation applications where the generated power and the efficiency ($\eta$) both scale with the temperature gradient (e.g., $\eta_{Carnot} = \Delta T/T_{hot}$). In sphere-plane geometries[19-21], distances as low as 20 nm[21] were achieved, which is close to the distance where the deep sub-wavelength regime occurs (typically 10-20 nm)[20]. However, this configuration allows near-field heat transfer only over a small area at the tip of the sphere, and is hence not practical for energy generation applications where the power is proportional to the area.

Here we show radiative heat transfer in the deep sub-wavelength regime between two parallel structures using precise positioning with integrated microelectromechanical actuators (MEMS). Our system relies on parallel nanobeams monolithically integrated with electrostatic comb drive actuators (Fig. 1). These actuators allow precise displacement control, limited in theory to sub-nm precision by Brownian motion and by actuation voltage uncertainty (see supplementary information S3). Their power consumption is also low (<30 pW in the ON state), more than three orders of magnitude lower than the typical heating power applied to the system.

We rely on the surface phonon-polariton resonance of SiC to create the surface waves responsible for near field heat transfer. Note that although there has been a lot of theory work on near-field heat transfer with thin SiC films[2,6,22-24], no experimental work has been reported. We use silicon carbide deposited by plasma enhanced chemical vapor deposition (PECVD) and annealed to create a microcrystalline ($\mu$-SiC) phase (see methods). We characterize the infrared permittivity ($\varepsilon$) of $\mu$-SiC for the first time and find it to be well described by a Lorentz-Drude relation (equation 1) that has a sharper infrared resonance than most commonly used $SiO_2$:

$$\varepsilon(\omega) = \varepsilon_\infty \left(1 + \frac{\omega_L^2 - \omega_T^2}{\omega_T^2 - \omega^2 - i\Gamma\omega}\right), \quad (1)$$

with $\omega_T$ = 789 cm$^{-1}$, $\omega_L$ = 956 cm$^{-1}$, $\varepsilon_\infty$ = 8, $\Gamma$ = 20 cm$^{-1}$ (see characterization and extended discussion in supplementary information S1.)

We simulate the heat transfer between the nanobeams using a Fourier modal method based on the fluctuational electrodynamics formalism[25,26] and we predict that the deep sub-wavelength regime occurs at distances <200 nm. The exact geometry of the nanobeams considered for this simulation is presented in supplementary information S2. For distances smaller than 200 nm (see Fig 1 b) the simulated results follow the typical $1/d^\alpha$ law that is characteristic of the deep sub-wavelength regime. The geometry dependent factor $\alpha$ = 1.75 in this case is intermediate between the parallel plate case ($\alpha$ = 2) and the sphere-plane case ($\alpha$ = 1.5). The simulations also show that heat transfer in the deep sub-wavelength regime is quasi-monochromatic (see Fig 1 c) and centered around the $\varepsilon(\omega)$ = -1 surface phonon resonance frequency of SiC (937 cm$^{-1}$).

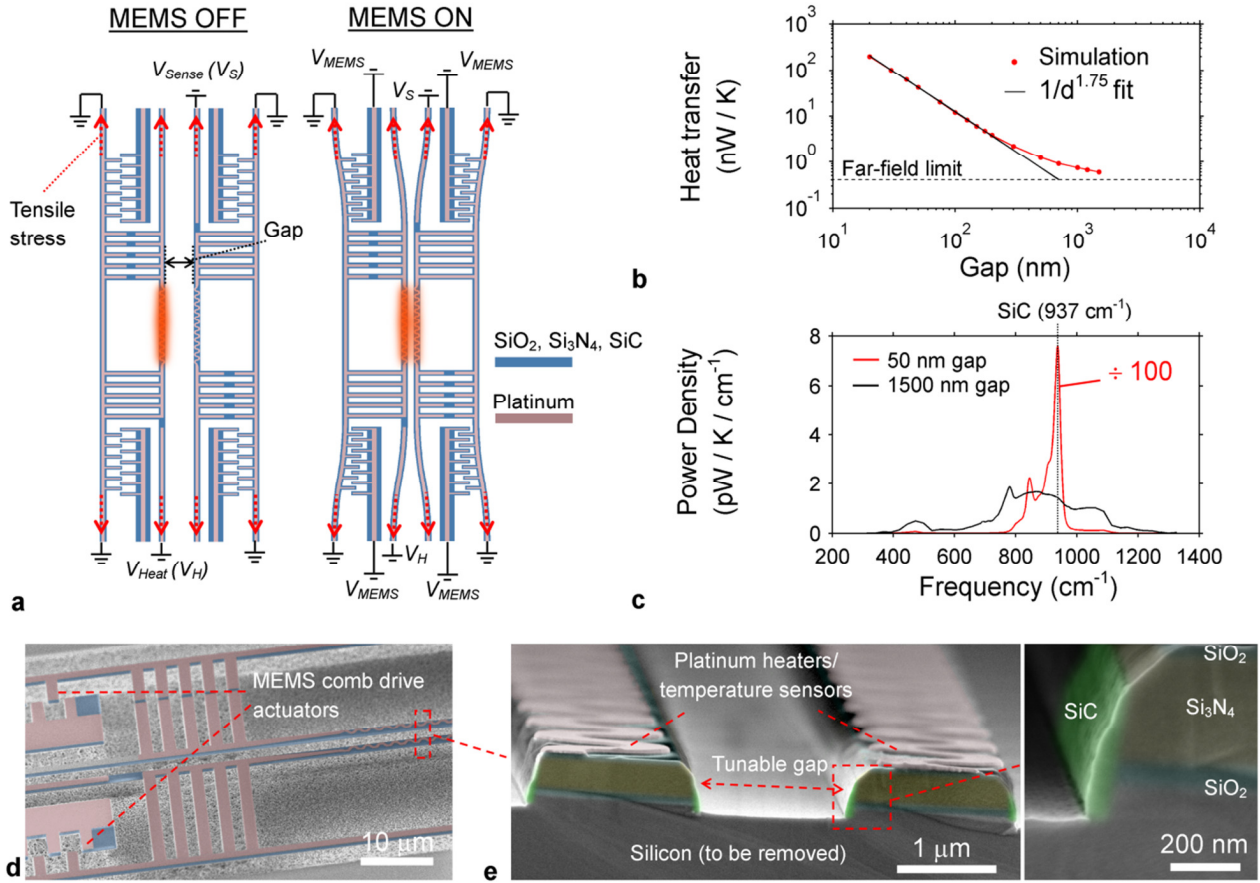

Figure 1 (a) Schematic of the device geometry (not to scale) and operation principle. (b) Simulated radiative heat transfer between two nanobeams as a function of the distance. The hot beam is set at $T = 10$ K above room temperature. The deep sub-wavelength regime occurs for beam separation below 200 nm, where the heat transfer begin to scale as $1/d^{1.75}$ (c) Simulated heat transfer spectrum between the nanobeams at two different separation. For $d = 50$ nm, the heat transfer is concentrated mainly around the SiC surface resonance (near $\omega = 937$ cm$^{-1}$). The secondary peak (at $\omega = 844$ cm$^{-1}$) results from a contribution of $Si_3N_4$ to the heat transfer. (d) False color scanning electron micrograph (SEM) of the device after structural release. (e) SEM false color cross section view of the nanobeams prior to structural release.

In order to achieve experimentally very small distances over large areas, it is critical that the two surfaces are completely parallel. Therefore no buckling should occur under changes of temperature. We prevent buckling by designing the beams such that they preserve the high tensile stress of silicon nitride (~900 MPa) after structural release (see Fig. 1 a). This stress prevents thermal bimorph effect from causing buckling that would catastrophically impact the minimum achievable distance in our system at high temperatures (see Fig. 2). Our design also suppresses stress induced deformation that prevented us from reaching the deep sub-wavelength regime with our previous platform[27].

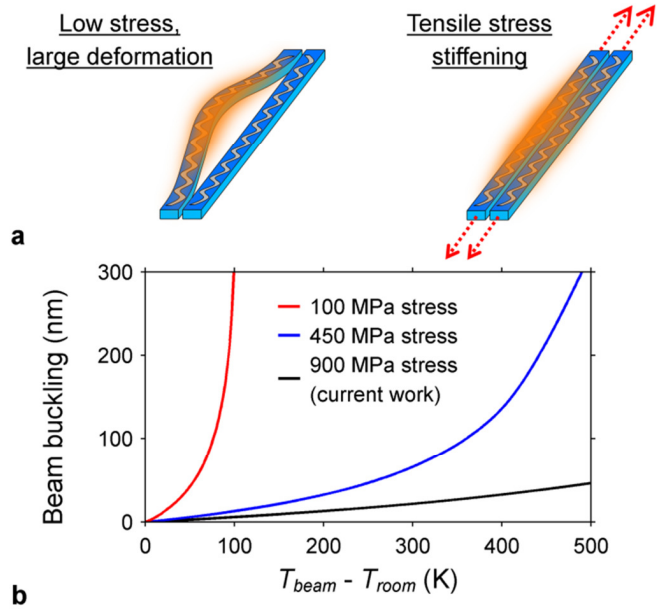

Figure 2 (a) Schematic of nanobeam buckling due to thermal bimorph effect. (b) Simulated buckling of a doubly clamped, 600 μm long, bi-material beam made of $Si_3N_4$ (300 nm thick) and Pt (50 nm thick).

In order to measure the heat transfer, the beams are brought together by sweeping the voltage supplied to the MEMS actuator (see $V_{MEMS}$ in Fig. 1 a) while the temperature of the heated beam ($T_{heat}$) and the sensing beam ($T_{sens}$) are measured (see Methods). The result of this scan is presented in Fig. 3 (b), where the inset shows the transition between near-field and contact regimes. The displacement of the nanobeams as a function of the MEMS voltage is also measured in a separate experiment (supplementary information S3) and presented in Fig 3 (a). The experimental data of Fig 3 is converted to normalized heat transfer power ($q$, in W/K) using

$$q = \frac{XP_{heat}T_{sens}}{(T_{heat}-T_{sens})(T_{heat}+XT_{sens})} \qquad (2)$$

where ($X = \sigma_{sens}/\sigma_{heat}$) is the ratio of the background heat conduction ($\sigma$) of the two nanobeams, and $P_{heat}$ is the supplied heating power. From the symmetry of the system, $X \approx 1$ for scans of relatively low temperature amplitude such as in Fig 3 (b).

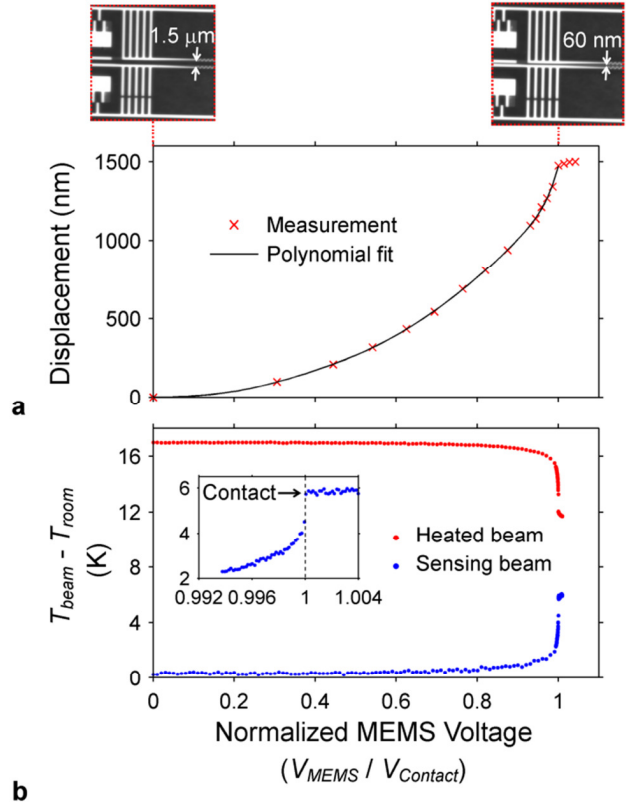

Figure 3 (a) Measured and fitted displacement of the nanobeams as a function of the voltage on the MEMS actuator. (b) Temperature of the heated beam ($T_{heat}$) and the sensing beam ($T_{sens}$) as a function of the voltage on the MEMS actuator.

The measured heat transfer agrees well with the simulated values. In Fig 4 (a), the experimental heat transfer power is plotted and compared with the simulated data (which is fitted vertically and horizontally, see methods). At gaps larger than ~150 nm, the experimental values are slightly larger than the simulated ones. This could be caused by larger infrared material absorption coefficient (of, e.g., $Si_3N_4$) than considered in the simulations, which would increase the contribution of propagating wave to the heat transfer. This increase causes the deep sub-wavelength regime to occur at <150 nm distances, rather than <200 nm in the simulated data (Fig 1 b). At gaps smaller than 150 nm, the experimental data matches the simulation more closely, and small deviations between simulation and experiments are visible only in the logarithmic scale inset in Fig. 4 (a). These are likely caused by the inability of the polynomial fit to perfectly match the experimental MEMS displacement in Fig 3 (a), or by slightly different experimental conditions between the MEMS displacement measurement (supplementary information S3) and the heat transfer measurement.

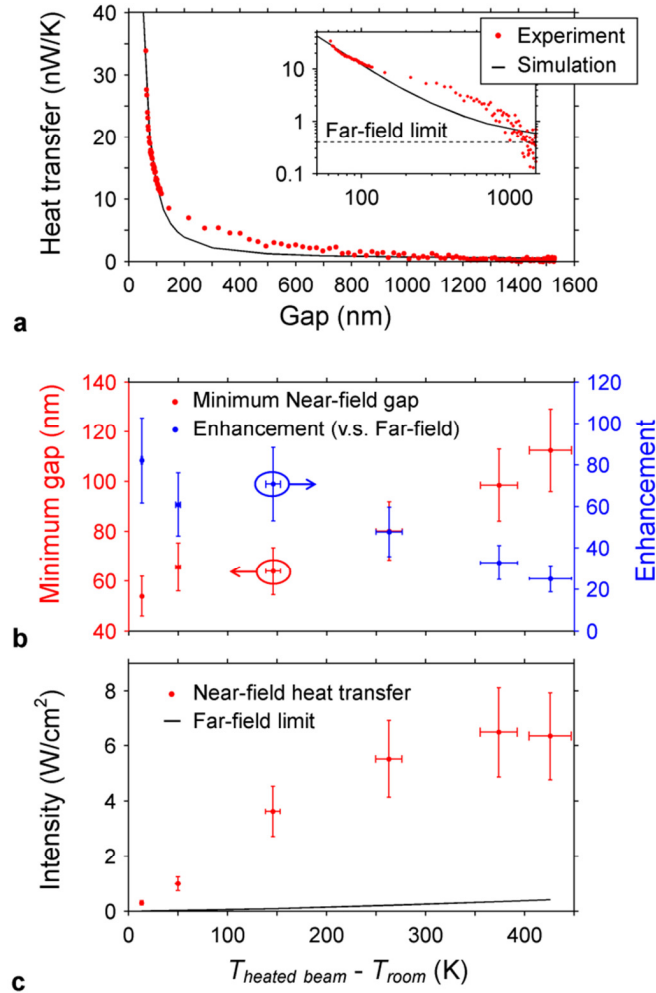

Figure 4 (a) Measured and simulated heat transfer between the nanobeams as a function of separation. The inset shows the same result, in logarithmic scale. (b) Minimum gap (i.e., before the beam come into contact) and maximum enhancement (relative to the far-field limit) achieved as a function of the heated beam temperature. This figure shows our ability to maintain the beam separation in the deep sub-wavelength regime (gap < 150 nm) for beam temperatures as high as 720 K. The error bars are calculated from the uncertainty on the thermal coefficient of resistance and the effective beam interaction length (see supplementary information S2). (c) Near-field heat transfer intensity (power per unit area) as a function of the temperature of the heated beam. The area is taken as the inner SiC surface of the nanobeams.

We achieve an 82× enhancement of heat transfer over the far-field limit and are able to maintain the nanobeams in the deep sub-wavelength regime for temperature gradients as high as 260 K. The highest heat transfer value achieved in Fig 4 (a) is 34 nW/K, which is 82× larger than the beam far-field radiation limit (0.42 nW/K). We define this limit as the maximum power that the SiC surface of the nanobeam can radiate to the far-field, at a given temperature (see Methods). This enhancement corresponds to a 54 nm gap between the nanobeams, which is well within the deep sub-wavelength regime (<150 nm). The measurement of Fig 4 (a) is repeated for different temperatures of the heated beam. In Fig 4 (b), the minimum achieved gap (i.e. the smallest near-field gap achieved before the beams come into contact with each other) is reported as a function of the heated beam temperature. The minimum achieved near-field gap is found to increase slightly with temperature. This is most likely due to thermal buckling that occurs despite the large stiffness of the nanobeams. The increase is indeed consistent with the simulated buckling presented in Fig 2. We reach the deep sub-wavelength regime (i.e. distance <150 nm) with heated beam temperatures as high as 720 K (i.e. 425 K above room temperature). In this case the temperature gradient between the beams is 260 K (see supplementary information S4). Such high thermal gradient translates to a high (>6 W/cm$^2$, see Fig 4 c) net energy flux which, for applications such as energy generation, is often more crucial than the relative enhancement itself.

We achieved the first demonstration of near-field radiative heat transfer between parallel objects in the deep sub-wavelength regime and at high temperature gradient. This nanoscale approach offers a clear path towards applications of near-field heat transfer such as near-field thermophotovolatic energy generation. We also note that our approach could be scaled up to a larger effective area by simply arraying several nanobeams (atop, e.g., a photovoltaic cell) and by individually controlling their out-of-plane displacement using MEMS actuators.

**Methods**

Microcrystalline Silicon Carbide ($\mu$-SiC) deposition
Amorphous silicon carbide is first deposited in by plasma enhanced chemical vapor deposition (PECVD) from silane ($SiH_4$) and methane ($CH_4$) gas precursors in a 1:20 ratio and with argon as a dilution gas. The film is then annealed in argon atmosphere at 1100 °C for 90 minutes to grow a microcrystalline phase[28]. The silane/methane concentration ratio was optimized by measuring the refractive index of the film by ellipsometry after the anneal process. The 1:20 ratio yields a 2.53 refractive index (at $\lambda$ = 1500 nm), which is very close to the theoretical value (2.57). Higher or lower ratio both lead to higher refractive index, which is consistent with the growth of silicon or carbon clusters during annealing of non-stoichiometric films.

Device fabrication
A $SiO_2/Si_3N_4/SiO_2$ film stack is first deposited on a conventional silicon wafer by low pressure chemical vapor deposition (LPCVD). The structure is then defined by deep-UV lithography and anisotropically etched in fluorine chemistry. SiC is then deposited over the structure by PECVD an annealed at 1100 °C to grow the $\mu$-SiC phase. SiC is anisotropically etched in fluorine chemistry, such that $\mu$-SiC remains only on the nanobeam sidewalls. Platinum heaters and metal contact are deposited be electron beam evaporation and lift-off using a chrome adhesion layer. The device is finally released by undercutting the silicon substrate in $XeF_2$ chemistry.

Experimental condition & procedure
The heat transfer experiments are performed at room temperature in a high vacuum (9 $\times 10^{-5}$ torr) electrical probe station. Electrical measurements are performed using an Agilent B1500 semiconductor parameter analyzer. The MEMS voltage ($V_{MEMS}$) is swept to bring the two nanobeams together while constant heating ($V_{Heat}$) and sensing ($V_{Sens}$) voltages are supplied to each of the two nanobeams. The time interval between each MEMS voltage increment (50 ms) is several times larger than the thermal response time of the system (7 ms), such that all measurements are in steady state. $V_{Sens}$ is kept much lower than $V_{Heat}$, such that the power supplied to the sensing beam is always at least 25 times lower than the power supplied to the heated beam ($P_{Heat}$). These constant voltages are also used to measure the temperature of the two beams, through the variation their electrical resistance (*R*) as $\Delta R/R = TCR \times \Delta T$, where $TCR$ = 0.00166 K$^{-1}$ is the measured temperature coefficient of resistance of platinum (see next section).

Temperature coefficient of resistance measurement
The temperature coefficient of resistance (*TCR*) of Platinum is measured by placing the device on a hot plate ramped from room temperature to 100 °C. We observe that the *TCR* differs greatly after the device is used at high temperature for the first time. For this reason, prior to *TCR* measurement, a heating voltage ($V_{Heat}$) 10% higher than the one used for the highest temperature point in Fig 4 (b, c) is supplied to the device under vacuum and held for 5 minutes until the current (and hence the electrical resistance) stabilizes. After this procedure, the device is placed on a hot plate ramped from room temperature to 100 °C and the resistance is measured at every 3 °C increment. We find that the thermal coefficient of resistance is 0.00125 K$^{-1}$ before the annealing procedure and $TCR$ = 0.00166 K$^{-1}$ after. We estimate a ±5% error on this value from the repeatability of the measurement.

Fitting procedure
In Fig 3 (a) the measured displacement as a function of the MEMS voltage is fitted using an 8th order polynomial. The fitted function is subsequently used to convert the MEMS voltage to displacement values. The polynomial order is chosen iteratively to be high enough to match the experimental points well, while being low enough to minimize spurious oscillation between the experimental points.

In Fig 4 (a) the experimental and theoretical data are fitted together in two different ways. Firstly, the experimental data is translated horizontally to account for the uncertainty on the initial gap between the nanobeams (i.e., the relative displacement of the beam is well known from Fig 3 (a), but the exact initial separation is unknown). The initial gap that best fit the experimental results is 1529 nm, close to the designed 1500 nm value. Secondly, the experimental data is translated vertically to account for parasitic heat conduction through the substrate. The translation that best fits the experiment is 1 nW/K, which is negligible compared to the achieved radiative heat achieved at the smallest gap (34 nW/K).

Far-field limit calculation
We define the far-field limit as the maximum power that can be radiated to the far-field by the inner Silicon Carbide part of the heated nanobeam. This calculation is performed using the Fourier modal method[25,26]. The far-field temperature in this simulation is set to room temperature (293 K).

**List of supplemental material**

S1 – Microcrystalline silicon carbide ($\mu$-SiC) infrared characterization
S2 – System dimensions
S3 – MEMS displacement measurement and theoretical precision limit
S4 – High temperature experimental data

**Acknowledgements**
The authors gratefully acknowledge support from DARPA for award FA8650-14-1-7406 supervised by Dr. Avram Bar-Cohen. This work made use of the Cornell Center for Materials Research Shared Facilities, which are supported through the NSF MRSEC program (DMR-1120296), and of the Cornell NanoScale Facility, a member of the National Nanotechnology Infrastructure Network, which is supported by the National Science Foundation (Grant ECCS-0335765). R.S.-G. held subsequent postdoctoral fellowships from the Fonds de recherche du Québec−Nature et Technologies, and from the Natural Sciences and Engineering Research Council of Canada (NSERC) during this work.

**Competing financial interest**
The authors declare no competing financial interests.

# Supplementary Information:
# Near-field radiative heat transfer between nanostructures in the deep sub-wavelength regime

Raphael St-Gelais,[1,2] Linxiao Zhu,[3] Shanhui Fan,[3] and Michal Lipson*[1,2]

[1]School of Electrical and Computer Engineering, Cornell University, Ithaca, New York 14853, United States
[2]Department of Electrical Engineering, Columbia University, New York, New York 10027, United States
[3]Ginzton Laboratory, Stanford University, Stanford, California 94305, United States
* e-mail: ml3745@columbia.edu

**S1 – Microcrystalline silicon carbide ($\mu$-SiC) infrared characterization**

We characterize the infrared permittivity of $\mu$-SiC by measuring the absorption of a 120 nm thick film deposited on a conventional silicon substrate. The film is deposited and annealed following the procedure described in Methods. We measure the absorption at 20 degrees grazing incidence using a Fourier transform infrared spectrometer (Bruker Hyperion). At grazing (rather than normal) incidence, measurements are sensitive to absorption from longitudinal optical phonons (LO), as well as from transverse optical (TO) phonons (which are usually the only ones visible in normal incidence spectra.) From this measurement (see Fig S1 a) we are able to fit the absorption of the film using conventional one-dimensional optical multilayer calculations and a Lorentz-Drude model for the complex permittivity of $\mu$-SiC,

$$\varepsilon(\omega) = \varepsilon_\infty \left(1 + \frac{\omega_L^2 - \omega_T^2}{\omega_T^2 - \omega^2 - i\Gamma\omega}\right), \tag{S1}$$

where $\omega_L$ and $\omega_T$ are the LO and TO phonon frequencies, $\varepsilon_\infty$ is the baseline permittivity, and $\Gamma$ is the phonon damping factor. For the simulation, we set the refractive index of the substrate to $n_{Si} = 3.42$, since silicon is transparent in this frequency range. The numbers that best fit the simulations are $\omega_T = 789$ cm$^{-1}$, $\omega_L = 956$ cm$^{-1}$, $\varepsilon_\infty = 8$, and a damping factor ($\Gamma = 20$ cm$^{-1}$) that is 5 times larger than what is reported for single crystal silicon carbide [S1]. Although the damping parameter is larger than for single crystal SiC, $\mu$-SiC can still be expected to yield narrower near-field heat transfer spectra than SiO$_2$, which is currently the most widely used material for near-field heat transfer experiments. In Fig S1 (b), we compare the local density of states (LDOS),

$$LDOS \propto \frac{\varepsilon''}{|1+\varepsilon|^2}, \tag{S2}$$

at surfaces of $\mu$-SiC and SiO$_2$, and find that $\mu$-SiC can be expected to yield a 5.5× narrower heat transfer frequency distribution than SiO$_2$. For this simulation, permittivity of SiO$_2$ is taken from [S2].

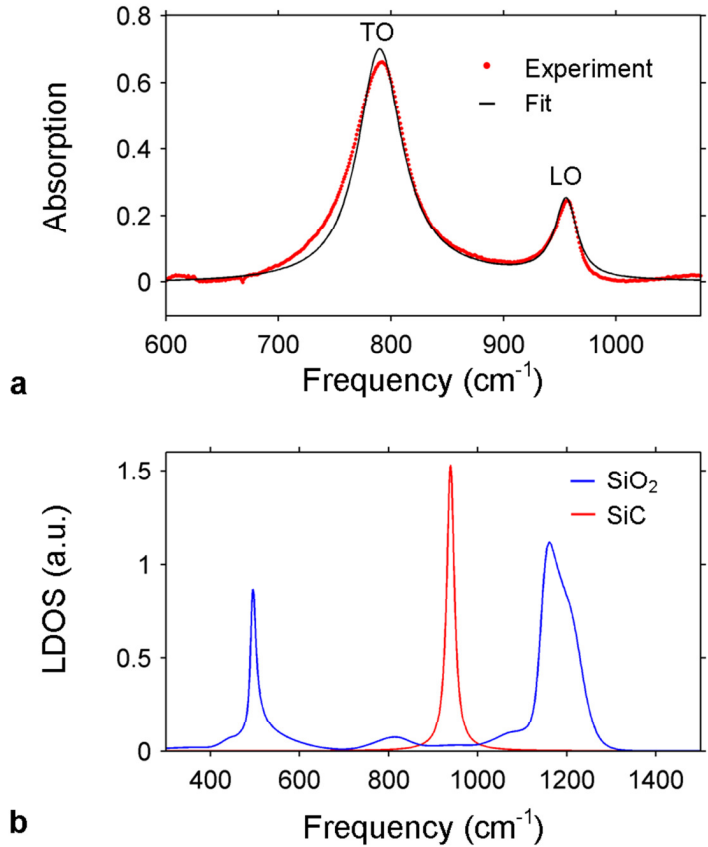

Figure S1. (a) Measured and fitted infrared absorption spectra of a 120 nm thin $\mu$-SiC film on silicon at 20 degree grazing incidence. (b) Comparison of the local density of states at the surfaces of $\mu$-SiC and $SiO_2$.

**S2 – System dimensions**

The cross section dimensions of the nanobeam considered for the near-field simulations are presented in Fig. S2. The dimensions of the MEMS are presented in Fig. S3. From the dimensions in Fig. S3, we estimate a 25% uncertainty on the effective beam interaction length due to the length difference between the beams parallel regions (200 μm) and the heater length (155 μm).

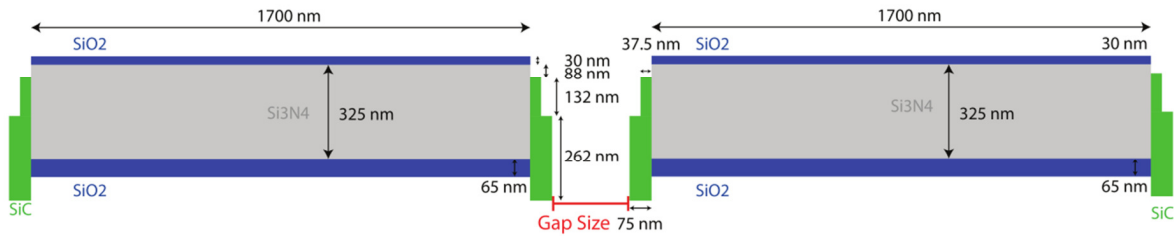

Figure S2. Beam cross-section considered for the simulations.

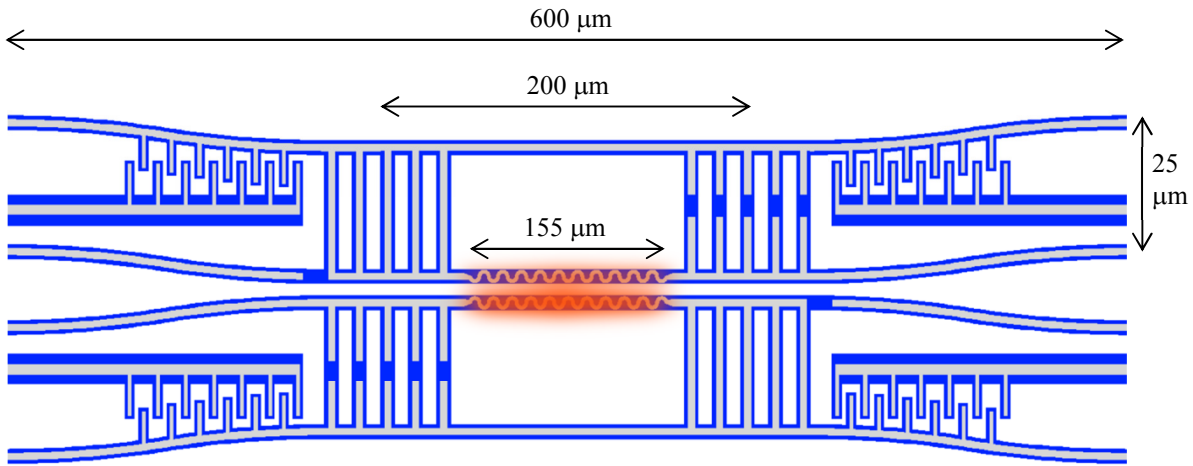

Figure S3. Main dimensions of the MEMS structure (drawing not to scale).

### S3 – MEMS displacement measurement and theoretical precision limit

The MEMS displacement as a function of the applied voltage is measured by placing the device on an electrical probe station equipped with a high magnification (50X) microscope. The displacement is measured as a function of the voltage applied on the MEMS actuator using an image treatment algorithm that fits Gaussian functions over each optical image of the nanobeams. The displacement is then evaluated by calculating the centroid of the fitted Gaussian functions.

The highest measured displacement sensitivity of the MEMS (see main paper, Fig 3 a) is 65 nm/V when the nanobeams are almost in contact (i.e., close to $V_{contact}$ = 155 V). The resolution of Agilent B1500 voltage source being 10 mV, this translates to a 0.65 nm resolution limit on the MEMS displacement. Fundamentally, the resolution on the MEMS displacement would eventually be limited by thermal mechanical motion ($x_{rms}$) given by [S3]:

$$x_{RMS} = \sqrt{\frac{k_B T}{K}}, \tag{S3}$$

where $K$ is the mechanical stiffness of the structures, which we estimate to 1-2 N/m. In this case, and considering the highest temperature used in our experiments, the amplitude of thermal mechanical vibration would be around 0.1 nm RMS.

### S4 – High temperature experimental data

Figure S4 presents the same temperature experimental data as in the main manuscript (Fig. 3 b), but for the highest temperature used in this work. At the smallest gap between the nanobeams (before contact), the thermal gradient is 261 K.

We note that in this case the two curves are asymmetric. This is caused by a higher background conduction ($\sigma$) of the heated beam compared to the background conduction of the sensing beam (i.e. $X = \sigma_{sens}/\sigma_{heat} \neq 1$, unlike for the low temperature scan in Fig 3 b). In the present case the ratio is $X = 0.28$. This is most likely caused by a higher far-field radiation of the heated MEMS compared to the sensing one when the temperature difference is very high.

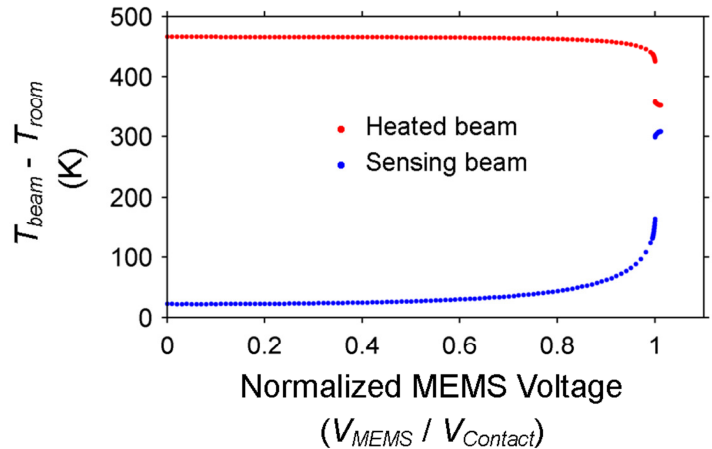

Figure S4. Measured temperature of the heated and sensing beam as a function of the MEMS voltage for the highest temperature scan performed in this work.